\documentclass[sigconf,authorversion,nonacm]{acmart}

\AtBeginDocument{%
  \providecommand\BibTeX{{%
    \normalfont B\kern-0.5em{\scshape i\kern-0.25em b}\kern-0.8em\TeX}}}

\graphicspath{{./images/}}

\usepackage{enumitem}
\usepackage{siunitx}
\usepackage{xspace}
\usepackage{multirow}
\usepackage{subfig}
\usepackage{afterpage}
\usepackage{svg}

\newcommand{\quotes}[1]{``#1''}
\newcommand{\schemefullname}{\emph{Low-Latency User-Anonymized DNS}\xspace}
\newcommand{\schemeacronym}{\texttt{LLUAD}\xspace}
\newcommand{\popularitylist}{\texttt{Popularity List}\xspace}
\newcommand{\loadbalancingpool}{\texttt{Load-Balancing Pool}\xspace}
\newcommand{\mixnetworkvotes}{\texttt{Voting Mix Network}\xspace}
\newcommand{\vote}{\texttt{Vote}\xspace}
\newcommand{\votes}{\texttt{Votes}\xspace}

\begin{document}
\sloppy
\title{LLUAD: Low-Latency User-Anonymized DNS}

\author{Philip Sjösvärd}
\affiliation{%
  \institution{Networked Systems Security group\\KTH Royal Institute of Technology}
  \city{Stockholm}
  \country{Sweden}}
\email{sjosv@kth.se}

\author{Hongyu Jin}
\affiliation{%
  \institution{Networked Systems Security group\\KTH Royal Institute of Technology}
  \city{Stockholm}
  \country{Sweden}}
\email{hongyuj@kth.se}

\author{Panos Papadimitratos}
\affiliation{%
  \institution{Networked Systems Security group\\KTH Royal Institute of Technology}
  \city{Stockholm}
  \country{Sweden}}
\email{papadim@kth.se}


\begin{abstract}
The Domain Name System (DNS) is involved in practically all web activity, translating easy-to-remember domain names into Internet Protocol (IP) addresses. Due to its central role on the Internet, DNS exposes user web activity in detail. The privacy challenge is honest-but-curious DNS servers/resolvers providing the translation/lookup service. In particular, with the majority of DNS queries handled by public DNS resolvers, the organizations running them can track, collect, and analyze massive user activity data. Existing solutions that encrypt DNS traffic between clients and resolvers are insufficient, as the resolver itself is the privacy threat. While DNS query relays separate duties among multiple entities, to limit the data accessible by each entity, they cannot prevent colluding entities from sharing user traffic logs. To achieve near-zero-trust DNS privacy compatible with the existing DNS infrastructure, we propose \schemeacronym: it locally stores a \popularitylist, the most popular DNS records, on user devices, formed in a privacy-preserving manner based on user interests. In this way, \schemeacronym can both improve privacy and reduce access times to web content. The \popularitylist is proactively retrieved from a (curious) public server that continually updates and refreshes the records based on user popularity votes, while efficiently broadcasting record updates/changes to adhere to aggressive load-balancing schemes (i.e., name servers actively load-balancing user connections by changing record IP addresses).
User votes are anonymized using a novel, efficient, and highly scalable client-driven \mixnetworkvotes{} -- with packet lengths independent of the number of hops, centrally enforced limit on number of votes cast per user, and robustness against poor client participation -- to ensure a geographically relevant and correctly/securely instantiated \popularitylist. We find that with a \num{25000} entries long \popularitylist, \schemeacronym provides both privacy-preserving and high performance DNS: this is due to the instant local (and anonymous) resolution of around \SI{94}{\percent} of queries based on the \popularitylist, with the few remaining queries using other privacy-preserving, but latency-costly, alternatives, such as querying a public resolver over a public anonymous network, e.g., Tor.
Beyond strong DNS privacy and low average lookup latency, \schemeacronym maintains network traffic overhead on par with widely deployed secure DNS protocols, with a memory/storage overhead of less than \SI{2}{\mega\byte}.
\end{abstract}




\maketitle

\section{Introduction}

The Domain Name System (DNS) is a key component of the Internet, translating domain names into Internet Protocol (IP) addresses and other types of related information \cite{dns_rfc, dns_privacy_recommendations_rfc, dns_resolver_cloudflare}. DNS resolvers respond to client queries with the sought information/translation, obtaining it from an internal cache, or by recursively querying a distributed hierarchical network of name servers to retrieve the desired DNS record from an authoritative name server \cite{dns_rfc}. To prevent external entities from manipulating DNS responses or observing user queries, existing secure DNS protocols enable resolvers to authenticate name server responses using Domain Name System Security Extensions (DNSSEC) \cite{dnssec}, and encrypt user-to-resolver traffic \cite{dot_rfc, doh_rfc, dnscrypt_protocol}. However, these efforts fail to protect user privacy: user DNS queries, directly reflecting their web activity -- revealing detailed history of visited domains, frequency of Internet activity, used Internet services, etc. -- remain exposed to the resolvers \cite{dns_privacy_recommendations_rfc}. This is particularly concerning given the rising popularity of public DNS resolvers; public resolvers account for nearly \SI{60}{\percent} of global DNS name server traffic, with Google DNS accounting for approximately \SI{30}{\percent} \cite{dns_resolver_relevance}. These popular public resolvers, often hosted by large corporations \cite{dns_resolver_google, dns_resolver_cloudflare, dns_resolver_relevance}, providing fast and reliable resolutions, raise concerns about the collection and potential misuse of sensitive user data. Notably, DNS provides particularly reliable data on user-accessed web resources, unlike the potentially unreliable inferences made, for example, by Internet Service Providers (ISPs) based on accessed IP addresses. That is, several domains can be hosted on the same IP address(es): see the widespread usage of Content Delivery Networks (CDNs).

DNS queries can be anonymized by introducing a relay server to separate the queried information from the sender identity \cite{odoh_rfc, anonymized_dnscrypt}, revealing the content of the query only to the resolver and the sender IP address only to the relay. However, the separated information can be trivially unified if the two entities collude.
In other words, relay server solutions rely on trust, without any further guarantees for privacy. Anonymous networks (e.g., Tor \cite{tor_paper}) provide a higher level of user anonymity at the expense of significant latency \cite{tor_performance, i2p_performance, anonymized_dnscrypt} to DNS query resolutions. To improve privacy without increasing latency, i.e., without sacrificing user experience, proposed solutions include: chaff queries, i.e., generating decoy traffic to obscure user queries/interests \cite{dns_range_queries, dns_broadcast_range_queries_mix_zones}, or pre-downloading a list of the most popular DNS records to anonymously resolve affected queries locally \cite{dns_broadcast_range_queries_mix_zones}. While the former suffers from high bandwidth overhead and potentially poor chaff quality \cite{dns_range_queries_pattern_analysis, dns_tracking}, the latter is a promising concept, as it eliminates exposure to the resolver for the majority of queries. However, several key questions remain: How do we instantiate and maintain the relevance of such a list of popular DNS records based on user interests/needs in a privacy-preserving manner? How do we achieve efficient incremental updates to the client-cached records, as they continuously change due to aggressive DNS load-balancing schemes? How do we design a more efficient format of the list to reduce its memory/storage overhead? And fundamentally in this context, how can we enable users to efficiently, securely, and anonymously vote on the content of such a list of popular DNS records?

To effectively address these challenges, we propose \schemefullname (\schemeacronym). \schemeacronym provides two methods to resolve DNS queries. First, the user attempts to resolve them locally using a pre-fetched \popularitylist of frequently resolved DNS records, maintained and incrementally updated by a public untrusted (honest-but-curious) server. In the event the resolution fails, i.e., the query response is unavailable in the \popularitylist, the user will resort to querying external DNS resolvers using other privacy-focused protocols, such as public mix networks (e.g., Tor).
To effectively instantiate \schemeacronym, we contribute:

\begin{enumerate}[leftmargin=20pt]
    \item A novel client-based \mixnetworkvotes where clients can \emph{efficiently} and \emph{anonymously} vote on geographically relevant DNS records, with packet lengths independent of the number of hops, and a centrally enforced maximum number of votes cast per user.
    \item A scheme to \emph{pre-fetch} and rotate the answers of actively load-balanced records to enable highly efficient incremental updates to the \popularitylist.
    \item Methods for reducing the memory/storage overhead of the \popularitylist.
\end{enumerate}

In the rest of this paper, which is based/expands on the early-stage ideas and preliminary results we presented in \cite{spw_paper}: Section~\ref{sec:background} provides an overview of the DNS along with privacy work relevant to the adversarial model and the problem definition in Section~\ref{sec:system_model_and_problem_statement}. Our scheme, \schemeacronym, is detailed in Section~\ref{sec:scheme}. We evaluate key metrics of \schemeacronym performance in Section~\ref{sec:results}, before concluding in Section~\ref{sec:conclusion}.


\section{Background and Related Work}
\label{sec:background}


\textbf{Domain Name System (DNS):}
DNS records are distributed across a hierarchical structure of independently managed DNS zones \cite{dns_rfc, dns_zones}, allowing organizations to host and manage their own domain names. Each zone includes name servers that respond to queries for its authorized domains. The hierarchical structure of DNS allows any zone to be found starting from a few root name servers. Users can traverse down the hierarchy of domains and subdomains by recursively querying name servers to find an authoritative one that holds the desired record \cite{dns_rfc}. For example, to resolve the Fully Qualifying Domain Name (FQDN) \textit{www.example.com.}, a user starts at the root zone (denoted by the trailing dot), then moves to the \textit{com} zone, and finally to the \textit{example.com} zone, where the records for \textit{www.example.com} are found.


\textit{Recursive DNS Resolvers:}
This need for multiple queries encourages the use of recursive DNS resolvers, which handle the repeated querying process and cache results to reduce latency and load on name servers. Public DNS resolvers, e.g., hosted by Google (8.8.8.8 and 8.8.4.4) \cite{dns_resolver_google} and Cloudflare (1.1.1.1 and 1.0.0.1) \cite{dns_resolver_cloudflare}, have become the standard for DNS resolutions \cite{dns_resolver_relevance} due to their convenience, low latency, and reliability. However, this means that users must trust these public resolvers, as they gain access to their DNS queries \cite{dns_privacy_recommendations_rfc}.

\label{sec:introduction_dns_load_balancing}

\textit{DNS Load-Balancing:}
DNS is inherently suitable for load-balancing client connections across servers. For instance, a DNS query response can include multiple IP addresses corresponding to different servers hosting the requested resource \cite{dns_load_balancing}. By alternating the order of these addresses for each client query, DNS resolvers effectively distribute connection requests among the available servers. Moreover, name servers can actively redirect users based on real-time server loads to \emph{actively} load-balance connections \cite{dns_load_balancing}.


\textbf{Encrypted DNS Queries:}
Since DNS queries are typically unencrypted, a recent push has been to develop protocols enabling encryption between users/clients and DNS resolvers. The first major protocol was DNSCrypt \cite{dnscrypt_protocol}, publicly implemented in a DNS resolver in 2011 \cite{dnscrypt_date}. In 2016, a new protocol called DNS over TLS (DoT) \cite{dot_rfc} was proposed. Unlike DNSCrypt, which uses a more proprietary encryption set-up supporting both UDP and TCP, DoT uses TLS. Another protocol, DNS over HTTPS (DoH), emerged in 2018 \cite{doh_rfc} and has seen wider adoption \cite{dns_encryption_rate}. DoH adds HTTP on top of TLS, simplifying the protocol's interface and improving privacy by making DNS queries less distinguishable from regular HTTPS traffic.


\textbf{DNS Relay Servers:}
However, encryption does not protect user privacy against curious DNS resolvers. Anonymized DNSCrypt \cite{anonymized_dnscrypt} and Oblivious DoH (ODoH) \cite{odoh_rfc} were introduced to improve privacy by routing the encrypted queries through a relay that masks the origin (IP address) from the DNS resolver. Due to end-to-end encryption between the client and the resolver, the relay cannot observe the queried resources. However, there is no guarantee that the relay and resolver do not collude to undermine privacy. To defend their privacy against curious colluding relay servers, users can send queries over the Tor network \cite{dns_over_tor, dnscrypt-proxy} using DoH over Tor (DoHoT), to anonymously query resolvers. However, Tor can significantly affect the user experience \cite{tor_performance, anonymized_dnscrypt}, introducing around \SI{400}{\milli\second} to over \SI{1}{\second} of latency per DNS query (see Section~\ref{sec:results_latencies}); especially considering that websites typically trigger multiple DNS queries \cite{dns_range_queries_pattern_analysis, dns_privacy_recommendations_rfc}.


\textbf{Chaff Queries:}
To avoid the latency incurred by mix networks (e.g., Tor), one could create an artificial anonymity set by introducing decoy DNS queries (chaff traffic) to obscure user queries \cite{dns_range_queries, dns_broadcast_range_queries_mix_zones}. However, generating plausible and realistic decoys is difficult \cite{dns_tracking, dns_range_queries_pattern_analysis}: to ensure not only consistency between the probability of selecting a specific chaff domain/record and its expected popularity, but also that external entities cannot trivially correlate subsequent queries to reveal the genuine ones, given that websites typically trigger a pattern of multiple DNS queries \cite{dns_range_queries_pattern_analysis}. Even if we presume realistic chaff query distributions and that obscuring query patterns are achievable, a vast pool of chaff domains to draw from would be needed. For instance, the database in \cite{domains_and_subdomains} tracks billions of global domains and subdomains, resulting in significant overhead for storing a sufficient number of domains. Moreover, the communication overhead scales linearly with the anonymity set size; emulating $N$ artificial users results in $N$ times greater computational and network overhead for Internet services and DNS infrastructure.


\textbf{Private Information Retrieval (PIR):}
A more efficient solution is for users to directly, yet anonymously, resolve DNS queries using Private Information Retrieval (PIR). In PIR, users query either multiple non-colluding servers for parts of the answer, or a single server with a database supporting such anonymous queries by fulfilling specific computational requirements \cite{pir_1, pir_2}. Thus, users can anonymously query the contents of the database without resorting to the trivial solution of downloading it in its entirety, an otherwise ideal solution if not for the overhead. However, a database of a large number of DNS records has a small size (in bytes) \cite{dns_broadcast_range_queries_mix_zones}, thus the ideal solution to download the entire database is not only feasible, but arguably a promising alternative.


\textbf{Pushing the Most Popular DNS Records:}
Distributing a list of the top \num{10000} most frequently resolved records \cite{dns_broadcast_range_queries_mix_zones} offers low-latency and privacy-preserving resolution for the majority of DNS queries, while the remaining queries to DNS servers are anonymized with a mix network \cite{dns_broadcast_range_queries_mix_zones}. The list provider is responsible for continually refreshing/re-querying the records whenever their Time-To-Lives (TTLs) expire, broadcasting any found updates (e.g., changes to IP addresses) to users \cite{dns_broadcast_range_queries_mix_zones}.

Although \cite{dns_broadcast_range_queries_mix_zones} discusses several strategies for generating and maintaining such a list, only a brief qualitative analysis is provided. Notably, \cite{dns_broadcast_range_queries_mix_zones} do not provide a rigorous solution for initiating and updating the list based on the interests of its users. Ensuring the list is relevant for the specific user base is important, as indicated by the difference in hit ratio between the optimal, regional, and global list in \cite{dns_broadcast_range_queries_mix_zones} (achieving \SI{83.9}{\percent}, \SI{68.7}{\percent}, and \SI{41.3}{\percent}, respectively). However, to ensure relevance, user interests/queries need to be regularly (if not continuously) securely and anonymously sampled. Furthermore, it is of great interest to reduce the overhead of such a scheme. In particular, reducing the significant network bandwidth overhead of incremental updates seen in \cite{dns_broadcast_range_queries_mix_zones}, caused by name servers that constantly change the IP addresses of DNS records to actively balance server loads.


\textbf{Sphinx Mix Network:}
Mix networks anonymize packet origins by mixing packets from multiple senders within a network of geographically distributed relay nodes \cite{loopix, sphinx}. This disassociates the incoming packets from the outgoing ones, making it difficult to trace their origins. To send packets, users encrypt them several times using different keys that only a predetermined path of specific nodes in the network can decrypt, ensuring that they are sufficiently disassociated from their origin.

A notable example relevant to this work is the Sphinx mix network, where each receiving node decrypts the message using a symmetric key derived using Elliptic-curve Diffie-Hellman between its private key and a (singular) public key element included in each message \cite{sphinx}. Instead of, for example, the Tor circuit-switched approach requiring recursive pre-established tunnels (tunnel to the third node is established through the tunnel to the second node, through the one to the first node, etc.) \cite{tor_paper}, the Sphinx mix network operates on a packet-switched model, with each message containing sufficient information to perform a key exchange. The Sphinx mix network has the ability to blind the public key element in each message to ensure that it does not remain static throughout the mix network path, in which case it would trivially expose the traveled path. \schemeacronym leverages the blinding technique in its \mixnetworkvotes, as detailed in Section~\ref{sec:scheme_blinding}.

\section{Problem Statement and System Model}
\label{sec:system_model_and_problem_statement}


We aim to minimize Internet user activity exposed to DNS system entities, which may collude and exchange information to link DNS queries to their senders' IP addresses. Motivated by the shortcomings of existing DNS privacy solutions, we aim for low communication overhead, preserving the lightweight nature of the DNS protocol, and especially low latency, an important aspect of the Internet/Web user experience. Similarly, the computational and memory/storage overhead of the scheme should be practically imperceptible by the user of a web browser, considering that this is the primary use case of such a privacy scheme. In summary, a significant set of often conflicting requirements that \schemeacronym must fulfill.


\textbf{System and Adversary Model:}
There are three different types of system entities: users/clients (primarily assumed to be devices from which users browse the web), a public \schemeacronym server central to our proposal, and existing servers within the public DNS infrastructure (public resolvers, name servers, query relays, etc.). Figure~\ref{fig:structure_entities} illustrates the roles of these entities as a user obtains the \popularitylist, explained in detail in Section~\ref{sec:scheme}. 

\afterpage{
    \begin{figure}[!t]
        \centering
        \includesvg[width=0.89\linewidth]{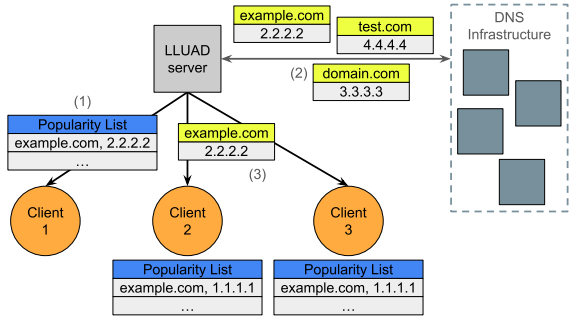}
        \caption{Example of a client downloading the \popularitylist (1), the \schemeacronym server re-querying the list's records as their TTLs continuously expire (2), and the \schemeacronym server pushing a discovered record update to connected clients (3).}
        \label{fig:structure_entities}
    \end{figure}
}

The public \schemeacronym server and the DNS servers/infrastructure, as well as ISPs, are considered honest-but-curious. Honest-but-curious entities provide their advertised/expected services, ensuring that the DNS protocol functions effectively and satisfies users. However, these entites are also curious about user activity in terms of DNS-resolved domains and records. They can drop or alter packets without denying DNS (e.g., occasionally dropped queries do not generally hinder DNS functionality). In contrast, users/clients are viewed not only as curious, but also as potentially malicious. Thus, clients might have malicious intentions to disrupt the protocol availability, by injecting, altering, or dropping packets. System entities may collude with each other to achieve their goals, assuming that it does not interfere with their overarching goals (e.g., an honest-but-curious server will not collude with a malicious client that is actively disrupting its service).

\section{Our Scheme}
\label{sec:scheme}

Central to our scheme is a \popularitylist locally stored on user devices, containing the $N_{popular}$ most popular DNS records in a specific area, geographic, or otherwise representative of its users. The size of $N_{popular}$ can vary per client based on available computational resources. This list acts as a local DNS resolver, either on the user device or within their personal network. Local DNS caches, such as those of web browsers, function as usual. That is, they are always prioritized over querying a resolver (including the \popularitylist). DNS queries not resolvable by the \popularitylist (the requested record is not in it) fall back to \schemeacronym-external privacy-preserving methods that maintain a high level of privacy, e.g., Tor, but at the cost of higher latency, offset by the primary reliance on the local \popularitylist. The format of the \popularitylist is detailed in Section~\ref{sec:scheme_popularity_list_format}.

The \schemeacronym server, as illustrated in Figure~\ref{fig:structure_entities}, manages and distributes the \popularitylist. Clients connect to the server and download the latest version of the list (step (1) in Figure~\ref{fig:structure_entities}). The \schemeacronym server continuously re-queries the public DNS infrastructure (name servers or public DNS resolvers) as the TTLs of records in the list expire (step (2)), broadcasting any found incremental updates to connected clients (step (3)). The \schemeacronym server and its accompanying \popularitylist emulate the role of a public DNS resolver. For example, name servers returning different results based on client-location would do so based on the location of the \schemeacronym server, much like they would with an intermediate public DNS resolver. Functionality such as the EDNS(0) Client Subnet Extension, which allows resolvers to forward the subnet of the querying user to allow the authoritative name server to return a geographically appropriate response \cite{edns_client_subnet}, is not supported due to its implications on privacy.

The relevance of the records in the \popularitylist is ensured by users (connected clients) continually casting anonymous \votes on records to include in the list. These \votes, reflecting user queries during the last $t_{refresh}$ seconds (saving resolved records as potential \votes), are submitted simultaneously by all clients at set $t_{refresh}$ intervals through a \mixnetworkvotes (detailed in Section~\ref{sec:scheme_mix_network}), optimized for our \schemeacronym scheme, allowing the \schemeacronym server to periodically refresh the \popularitylist. The \votes only specify which DNS records to include, not their corresponding translations since the \schemeacronym server performs the lookups. This prevents any potentially malicious client from poisoning records in the list.

\begin{figure}[!t]
    \centering
    \includesvg[width=0.89\linewidth]{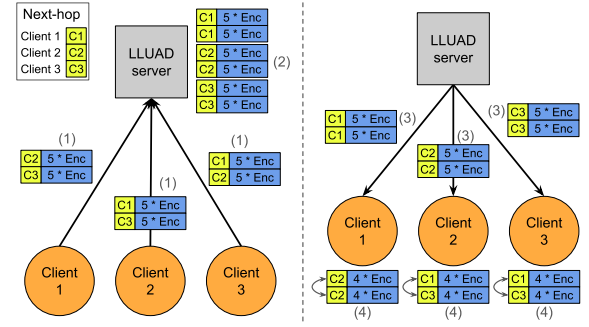}
    \caption{Illustration of the \schemeacronym server relaying batches of encrypted \votes between clients that decrypt the top encryption layer and shuffle their order.}
    \label{fig:shuffling_votes}
\end{figure}

As shown in Figure~\ref{fig:shuffling_votes}, our \mixnetworkvotes does not use third-party servers to mix \votes. Instead, connected clients act as mix nodes that collectively anonymize the \votes. Another distinct feature of the \mixnetworkvotes is that the \schemeacronym server relays packets (\votes) between mix nodes (clients), a more efficient and implementable approach (see Section~\ref{sec:scheme_mix_network}) compared to nodes directly communicating. The process of the \mixnetworkvotes, as illustrated by the enumerated steps in Figure~\ref{fig:shuffling_votes}, is as follows:

\begin{enumerate}
    \item All clients submit their \votes simultaneously to the \schemeacronym server. The \votes are iteratively encrypted using the keys of, in this example, five other mix nodes (clients) in the network, meaning they need to be received and processed by a sequence of five predetermined mix nodes (not counting the intermediate hops via the \schemeacronym server) before their content is revealed.
    \item The \schemeacronym server sorts the received \votes into batches based on their next-hop destination.
    \item The \schemeacronym server redistributes the batches of \votes to their corresponding client mix nodes.
    \item After receiving a batch of \votes, each client decrypts each top layer of encryptions and randomly shuffles the order of their \votes, obscuring their origins among each other.
\end{enumerate}

The process is repeated until all \votes are fully decrypted, allowing the server to evaluate the current popularity of records and update the \popularitylist accordingly. The voting mechanism is not latency sensitive. Therefore, the system can accommodate longer delays between each shuffling round to support users with high latency.

To mitigate Sybil attacks, where attackers connect with a large number of identities to obtain greater influence on the \popularitylist or undermine the anonymization of the \mixnetworkvotes, users must register in advance to an external Public Key Infrastructure (PKI) and authenticate with their obtained credentials (certificate) when connecting to the \schemeacronym server. Note that, similarly to anonymous networks, e.g., Tor, where ISPs can trivially observe users connecting to Tor entry nodes, \schemeacronym does not anonymize the identity of connected users to the \schemeacronym server, but rather their activity. The registration process can require a phone number, etc., to thwart large-scale Sybil attacks. To ensure the \schemeacronym server does not perform similar attacks, users can validate the identity of peer clients via the certificates obtained from the external PKI during the registration. Specifically, these certificates are used to verify the public keys of other clients in the \mixnetworkvotes (detailed in Section~\ref{sec:scheme_packet_format_overview}). Privacy can be further improved by using ephemeral pseudonymous certificates (used for vehicular communication settings \cite{vpki_1, vpki_2}) to hide user/device identifiable information. \footnotemark[1]

\footnotetext[1]{The ramifications for different approaches of registering users and distributing public keys/certificates cannot be fully addressed in this paper and are generally orthogonal to the privacy and quality of DNS resolutions. Thus, we will not elaborate further.}

\subsection{Construction of the Popularity List}
\label{sec:scheme_popularity_list_format}

The \popularitylist stored by clients contains only the domain, record type, and answer of the top $N_{popular}$ DNS records, excluding the root domain. Thus, redundant data such as DNS class and TTLs are omitted (the latter is tracked by the \schemeacronym server, which keeps the records up-to-date). Apart from the top $N_{popular}$ records, the \popularitylist also includes additional records that at least one of the top records redirects to using so-called CNAME records (records that redirect/point to another domain). The \popularitylist uses a hierarchical structure to minimize its overall size, avoiding duplicate data such as repeating the Top-Level Domain (TLD) \quotes{.com} after each domain entry. For instance, the \popularitylist in Figure~\ref{fig:popular_list_structure} contains four \quotes{.com} domains, yet only stores \quotes{com} once, as all four domains, due to their position in the tree, implicitly belong to the TLD. Each record answer in the tree (indicated in blue in Figure~\ref{fig:popular_list_structure}) is compacted, only containing the record type identifier and the corresponding answer (e.g., an IP address). Users query the local \popularitylist service using the DNS protocol, where the service searches the structure and responds with a correctly formatted DNS response. If the sought record is not found in the local list, the query is forwarded to an external privacy-preserving but latency-costly DNS service (see Section~\ref{sec:scheme_external_resolutions}).

To resolve \textit{mail.internal.example.com} in Figure~\ref{fig:popular_list_structure}, the service starts at \textit{com} and goes directly to the \textit{example} section. Since \textit{www} is not the correct subdomain, the first subdomain is skipped, finding the sought \textit{mail.internal}. In this example, \textit{mail} and \textit{internal} are combined to reduce computational and memory overhead, as the latter has no answers and only one subdomain (\textit{mail}) directly below it. CNAME records do not store their referenced domain in cleartext. Instead, they contain pointers to relevant domain labels in the \popularitylist to reduce overhead. For example, the CNAME record of \textit{www.example.com} in Figure~\ref{fig:popular_list_structure} could redirect to \textit{example.com} by pointing to the headers of \textit{com} and \textit{example}.

\subsubsection{Actively Load-Balanced Records}
\label{sec:scheme_active_load_balancing}

\begin{figure}[!t]
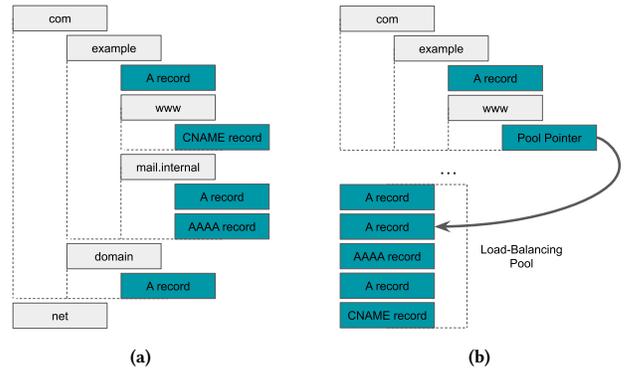

    \centering
    \subfloat[]{
        \label{fig:popular_list_structure}
        \includesvg[width=0.45\linewidth]{images/scheme/popularity_list_structure.svg}
    }
    \hfill
    \subfloat[]{
        \label{fig:load_balancing_pool_structure}
        \includesvg[width=0.45\linewidth]{images/scheme/load_balancing_pool_structure.svg}
    }
    \caption{(a) Example of a \popularitylist, (b) with an appended \loadbalancingpool containing all discovered answers of all actively load-balanced records.}
    \label{fig:structure}
\end{figure}

As described in Section~\ref{sec:introduction_dns_load_balancing}, DNS records can be actively load-balanced to dynamically distribute connections across servers. These records often have TTLs as short as \SI{30}{\second} and change with almost every query. This would result in substantial network traffic as the \schemeacronym server would broadcast each incremental update to the \popularitylist. To reduce network overhead, clients store all potential answers for all actively load-balanced domains in a shared \loadbalancingpool at the end of the \popularitylist (e.g., all IP address answers observed by the server as it re-queries these records), as illustrated in Figure~\ref{fig:load_balancing_pool_structure}, omitting duplicate answers. The affected entries in the \popularitylist tree structure point to the byte location of their corresponding answer in the pool.

Prefetching all potential answers allows the \schemeacronym server to efficiently update a record answer in the \popularitylist by sending two integers: the number of the \popularitylist entry to update (counting only actively load-balanced ones) and how many answers the pool pointer should be offset by, rather than sending the full record and corresponding answer, substantially reducing communication overhead. Non-load-balanced domains store their answers directly in the tree structure of the \popularitylist, as exemplified by \textit{example.com} in Figure~\ref{fig:load_balancing_pool_structure}. The answers of actively load-balanced domains are added in alphabetical order to the \loadbalancingpool based on (i) domain name, (ii) numerical record type, and (iii) answer content, omitting those already added.

To reduce memory/storage and bandwidth overhead, clients receive only one randomly selected answer per record type and domain from the \schemeacronym server when multiple answers are available. For example, if a DNS response (from an authoritative name server) contains three different A record answers (IPv4 addresses), the \schemeacronym server only distributes one randomly chosen address to each client. This preserves passive load balancing (assuming answers are distributed uniformly) while reducing overhead. Since there is only one active/stored answer per record, incremental updates to actively load-balanced records only require offsetting a single pool pointer instead of multiple, further minimizing network overhead.

To avoid the otherwise large overhead of packet headers, multiple incremental updates to actively load-balanced records are concatenated into a single transmitted message, sent at most once per minute by default. As a result, records with a TTL shorter than one minute may not be fully adhered to. However, the Chromium DNS cache (the basis of most web browsers) caches records for exactly one minute, regardless of the specified TTL \cite{chromium_cache_ttl_1, chromium_cache_ttl_2}. Thus, enforcing a one-minute minimum TTL has no practical impact on the DNS protocol.

\subsubsection{Number of Votes to Contribute}
\label{sec:scheme_tuning_parameters}

Two primary parameters control the number of \votes each client can/will contribute per voting round, determining the communication and computational overhead, and any client's influence on the \popularitylist. Moreover, the difficulty of reconstructing individual user profiles from the collected pool of \votes increases if users refrain from voting on most of their queries performed during the last $t_{refresh}$ period. The configurable parameters are:

\begin{enumerate}
    \item \textbf{Voting-rate:} The probability that a client will submit any performed query as a \vote to the \schemeacronym server.
    \item \textbf{Maximum \votes per voting-round:} The upper limit of number of cast \votes per $t_{refresh}$ period per client.
\end{enumerate}

Clients do not include how many times they have queried the record of each \vote, as it is difficult to enforce the correctness of said value. Furthermore, this way, the \popularitylist reflects the popularity of DNS records of a general population rather than the niche interests of highly active individual users. Although the \schemeacronym scheme does not prevent clients from maliciously submitting multiple \votes for the same DNS record, the impact of such behavior is thwarted by restricting the maximum number of \votes per client and refresh period (parameter (2) above). The details of how our scheme limits the number of anonymous \votes per user are given in Section~\ref{sec:scheme_mix_network}.

\subsubsection{Impact of Each Voting Round}

The user-cast \votes determine which $N_{popular}$ records are included in the \popularitylist. Each record is ranked according to its weighted sum of received \votes from each voting round (at the end of each $t_{refresh}$ period). The weighted sum of the latest $m$-th system-wide voting round is
\begin{equation}
    N_{weighted_m} = a \cdot N_{occurences_m} + (1 - a) \cdot N_{weighted_{m-1}} \,,
    \label{equ:weight}
\end{equation}
where $a$ is the weight for the latest voting round, $N_{occurences_m}$ is the number of \votes received for the record during voting round $m$, and $N_{weighted_{m-1}}$ is the weighted sum from all previous rounds. Initially, $N_{weighted_0} = 0$. The weight $a$ is empirically chosen to maximize the \popularitylist hit ratio, while balancing the volume of updates and the responsiveness to changes in popularity.

After updating record rankings at the end of a $t_{refresh}$ period, the \schemeacronym server sends messages to connected clients to inform which records/answers to remove/add to the \popularitylist and its \loadbalancingpool to update them accordingly. Messages removing records/answers are compacted by referencing domains in the same manner as \popularitylist CNAME entries.

\subsection{Mix Network for Vote Anonymization}
\label{sec:scheme_mix_network}

The simplest way for clients to vote anonymously would be through a public anonymous network (e.g., Tor). However, such third-party networks would prevent the \schemeacronym server from limiting the number of \votes cast per user, allowing attacks on the voting system, diluting and effectively poisoning the \popularitylist with unpopular records, to lower its hit ratio. Furthermore, to maintain unlinkability between \votes, each user would have to establish a new circuit (series of tunnels) for every \vote, making this approach inefficient. We address the above issues with our specialized \mixnetworkvotes.

\textbf{Objectives:} The \mixnetworkvotes must ensure that individual \votes are anonymous and that no one, including the \schemeacronym server, can link multiple \votes to the same (anonymous) user profile. Furthermore, \schemeacronym should not rely on volunteering third-party hosted mix node servers, or assume full trustworthiness of users. Additionally, it must function independently of firewalls and Network Address Translations (NATs) and impose minimal bandwidth overhead, comparable to the lightweight DNS protocol. Recall also that \schemeacronym needs to restrict the number of \votes per user (to safeguard the \popularitylist).

Having all clients simultaneously submit their \votes to the \mixnetworkvotes results in the maximum achievable anonymity set of potential packet/\vote origins. Additionally, each user \vote is sent through a unique \mixnetworkvotes path (a unique pseudo-random sequence of client mix nodes) to reduce \vote linkability between \votes from the same user. A compact packet format is used to minimize the bandwidth overhead typically associated with mix networks. Specifically, the lengths of the \vote packets are independent of the number of mix node hops and do not require pre-established tunnels. To ensure minimum leakage of user information, the number of \votes cast (per voting round) by each client is obscured. In particular, clients with fewer \votes than the maximum allowed limit will submit empty ones to the \mixnetworkvotes to fulfill the quota.

By routing all packets through the \schemeacronym server, the system improves efficiency by reusing existing connections, bypassing the need for clients to continuously open new connections to distribute the \votes, all with unique next-hop addresses. Reusing the server connection also avoids firewall and NAT implications. In addition, multiple \votes can be concatenated in the same IP packet to reduce the overhead of packet headers. Lastly, the server can enforce a limit on the number of \votes sent by each client, as it controls all flows in the \mixnetworkvotes. The synchronization of mixing/shuffling rounds prevents the server from analyzing the packet flows further, thwarting any risks of tracing the origin of \votes, an otherwise considerable issue with a single entity observing all mix node flows in a mix network.

\subsubsection{Packet Format}
\label{sec:scheme_packet_format_overview}

Each \vote passes through a set of client mix nodes and exits at the \schemeacronym server. The number of these mix nodes, $N_{shuffle}$, determines the number of rounds that \votes are shuffled. However, the number of stacked encryptions is $N_{shuffle}+1$, as the final encryption is removed by the \schemeacronym server. Each of the $N_{shuffle}+1$ target destinations has index $i = \{1, 2, ..., N_{shuffle}+1\}$.

Each \vote sent through the \mixnetworkvotes consists of three components: a public key element, $p_i$, a next-hop hash, $h_i$, and a payload, $d_i$. As indicated by the index of each component, all fields are transformed at each traversed node to maintain the integrity and anonymity of the \vote. If each element did not change, it would be trivial for an observing party (such as the \schemeacronym server) to map incoming and outgoing \votes at each shuffling client and deduce the traveled path, thus rendering the packet mixing obsolete. $Node_i$ derives $p_{i+1}$, $h_{i+1}$, and $d_{i+1}$ from $p_i$, $h_i$, and $d_i$ and sends them to the \schemeacronym server, which forwards them to $Node_{i+1}$.

\textbf{Public key element:} a \num{32}-byte elliptic-curve public key element \cite{sphinx} used by a receiving $Node_i$ to derive, when performing Diffie-Hellman with its private key, the symmetric key $s_i$ used for the current top encryption layer of a \vote packet.

\textbf{Next-hop hash:} a \num{16}-byte field functioning as an address, mapping to the identity of the current next-hop client mix node using a function similar to Distributed Hash Tables (DHTs). For example, the hash value modulo the number of client mix nodes, resulting in $h_i$ to map to the address of $Node_i$, $h_{i+1}$ to $Node_{i+1}$, etc.

\textbf{Payload:} a \num{32}-byte field containing the record the user is voting on, the first byte being a general flag field mentioned in Section~\ref{sec:scheme_payload_format}. The original payload $d$ is iteratively encrypted with symmetric keys $\{s_{N_{shuffle}+1}, s_{N_{shuffle}}, ..., s_i, s_{i-1}, ..., s_1\}$ (in order), where the original sender and $Node_i$ share the symmetric key $s_i$ ($Node_{N_{shuffle}+1}$ is the \schemeacronym server). Thus, the payload $d_i$ received by $Node_i$ is encrypted with $\{s_{N_{shuffle}+1}, s_{N_{shuffle}}, ..., s_{i+1}, s_i\}$.

\subsubsection{Next-Hop Hash}
\label{sec:scheme_next_hop_hash}

The next-hop address of each node in a mix network path is typically appended at each encryption stage, which can be later obtained by removing the top encryption layer. However, appending multiple addresses would add noticable overhead to the \schemeacronym voting system. Instead, each \vote packet uses a fixed-length pseudo-random hash that maps to a next-hop address and is rehashed after each mix node to reflect the next destination. $Node_i$ recalculates the received hash $h_i$ with
\begin{equation}
    h_{i+1} = Hash(h_i, s_i, t_{timestamp}) \,,
\end{equation}
where $t_{timestamp}$ is a system-wide variable representing the current voting round, ensuring a unique next-hop sequence, and $h_1$ is initialized as a random value. While $t_{timestamp}$ and $h_i$ are known/visible to observers, $s_i$ is only known by the original sender and $Node_i$. Thus, only the original sender and $Node_i$ can create $h_{i+1}$, where only the sender can derive the entire sequence of mix nodes with all required symmetric keys $\{s_1, s_2, ..., s_{N_{shuffle}+1}\}$.

The next-hop hash does not explicitly indicate when the packet should exit the \mixnetworkvotes. However, the \schemeacronym server that orchestrates the shuffling rounds can trivially track the current round number and update the \popularitylist with the \votes in the final round. Note that consequently, all clients must follow the same system-wide $N_{shuffle}$.

\subsubsection{Blinding the Public Key Element}
\label{sec:scheme_blinding}

Ensuring that all fields of the \vote packets continuously change each hop is generally straightforward. For instance, the next-hop hash is automatically refreshed into a new seemingly random value, and so is the payload after each decryption. However, the public key element $p_i$ used to derive $s_i$ remains inherently static. Thus, \schemeacronym uses a technique found in the Sphinx mix network \cite{sphinx}, where the public key element is blinded at each hop, transforming it into another public key element. The transformation involves the shared key, $s_i$, which means that only the original sender and the specific mix node know the mapping between the incoming and the outgoing public key elements ($p_i$ and $p_{i+1}$). $Node_i$ blinds $p_i$ with
\begin{equation}
    p_{i+1} = Blind(p_i, b_i) \,,
\end{equation}
where
\begin{equation}
    b_i = Hash(p_i, s_i)
    \label{equ:blinding}
\end{equation}
is the blinding factor applied by $Node_i$. The symmetric key $s_i$ shared between the original sender and $Node_i$ is derived by the latter using Diffie-Hellman (DH):
\begin{equation}
    s_i = DH(p_i, k_{priv_i}) \,,
\end{equation}
where $k_{priv_i}$ is the private key of $Node_i$. Much like how the original sender can create the entire chain of next-hop hashes, $h_i$, for all $i = \{1, 2, ..., N_{shuffle}+1\}$ during the creation of the mix network packet, they can also generate $p_i$ for all $i$. Table~\ref{tab:scheme_vote_components} shows all the steps for generating the original \vote packet, where the final step is to iteratively encrypt the payload, $d$, with all derived symmetric keys.

\begin{table*}[!t]
  \begin{center}
    \caption{Calculated elements, in order left to right and top to bottom, to create a \mixnetworkvotes \vote packet.}
    \label{tab:scheme_vote_components}
    \begin{tabular}{|l|l|l|l|l|}
      \hline
      \textbf{Step} & \textbf{(1) Next-hop hash} & \textbf{(2) Public key} & \textbf{(3) Symmetric key} & \textbf{(4) Blinding} \\ \hline
      \num{1} & $h_1 = Random\ hash\ value$ & $p_1 = Gen_{public}(k_{priv_0})$ & $s_1 = DH(k_{priv_0}, k_{pub_1})$ & $b_1 = Hash(p_1, s_1)$ \\ \hline
      \num{2} & $h_2 = Hash(h_1, s_1, t_{timestamp})$ & $p_2 = Blind(p_1, b_1)$ & $s_2 = DH(k_{priv_0}, k_{pub_2}, b_1)$ & $b_2 = Hash(p_2, s_2)$ \\ \hline
      \multicolumn{5}{c}{\textit{...}} \\ \hline
      $i\ (\leq N_{shuffle})$ & $h_i = Hash(h_{i-1}, s_{i-1}, t_{timestamp})$ & $p_i = Blind(p_{i-1}, b_{i-1})$ & $s_i = DH(k_{priv_0}, k_{pub_i}, b_1, ..., b_{i-1})$ & $b_i = Hash(p_i, s_i)$ \\ \hline
      \multicolumn{5}{c}{\textit{...}} \\ \hline
      $N_{shuffle} + 1$ & \multicolumn{4}{|l|}{$d_1 = Enc(Enc(... (Enc(Enc(d, s_{N_{shuffle}+1}), s_{N_{shuffle}}), ...), s_2), s_1)$} \\ \hline
    \end{tabular}
  \end{center}
\end{table*}

\subsubsection{Payload Format}
\label{sec:scheme_payload_format}

Each \vote includes the type and domain of the corresponding record, where the latter can vary significantly in length. Since length is a distinct feature of individual packets, the \votes would be traceable throughout the \mixnetworkvotes. Thus, all packets are padded to a uniform length. A fixed number of bytes is allocated for the record type and domain name; specifically, the \num{31} last bytes of the payload $d$. If the combination of record type and domain exceeds this length, it is hashed to fit within the limit. The \schemeacronym server can interpret the hash if it has previous knowledge of the record.

While the server can build a database of known DNS records and their corresponding hashes over time, it must be able to identify records on their first encounter. The server could use an external service that provides practically all public domains \cite{domains_and_subdomains}. Nevertheless, to allow \schemeacronym to function independently, the server can request assistance from its clients. The server can broadcast a message to indirectly request the client who submitted the hash to anonymously provide the record in cleartext during the next voting round. Anonymity/Unlinkability is maintained as the client splits the cleartext record across several \vote packets, using the leading flag and a random ID number occupying the first two bytes of the \num{31} in the payload to help the server reconstruct the otherwise disassociated messages as they exit the \mixnetworkvotes.

\subsubsection{Tracking Other Clients}

Each client must pre-fetch the public keys of all clients in the paths of its \votes. Since mix node paths are randomly selected using the pseudorandom next-hop hashes (can at most re-roll the sequence by selecting a new initial next-hop hash $h_1$ to build the sequence from), this requires clients to store the public keys of all other clients in the system, leading to significant storage overhead in the case of many users. To reduce public key storage overhead, the shuffling role can be assigned to a subset of randomly chosen users, potentially rotating the assigned roles over time to distribute overhead and trust in any one entity. The additional overhead on these users is not of major concern, as the overhead of shuffling \votes is only a small part of the total overhead of \schemeacronym. To counteract collusion and ensure the integrity of the \mixnetworkvotes, the \schemeacronym server must not dictate the assignment of these users. Note that since the \schemeacronym server redirects \votes based on next-hop hashes, clients do not need to store the IP addresses of other clients.

As user devices could go offline at any moment, all participants must know which clients (assigned the shuffling role) are active and available to mix \votes. Thus, the \schemeacronym server initiates each voting round by broadcasting a message containing a bit stream with each bit representing the availability/online status of each client. The first bit implies whether the first shuffling client (first public key) should be considered, the second bit the second client (second public key), etc. This also allows users to assess the current size of the \mixnetworkvotes anonymity set.

\subsubsection{Vote Acknowledgments}
\label{sec:scheme_reply}

While all \votes can be anonymously contributed to the server through the \mixnetworkvotes, malicious actors could covertly perform several attacks. For example, a client participating in shuffling \votes could add, remove, or replace them. Although the \schemeacronym server can observe discrepancies in the number of sent and received \votes, thus revealing clients adding or dropping \votes, a client could still corrupt or replace \votes to influence the \popularitylist.
To detect whether a \vote was corrupted or replaced, the \schemeacronym server sends an authenticated acknowledgment for each received \vote, allowing users to confirm that the server correctly received each of their \votes, and be notified of any malicious behavior within each corresponding \mixnetworkvotes path. Users may report detected misbehavior (reporting the selected path of clients) to the \schemeacronym server, which can identify the malicious client by correlating sufficiently many reports. The acknowledgments are anonymously sent to the original senders by shuffling them in synchronized rounds similar to the \votes. Specifically, all acknowledgments traverse the reverse path of their corresponding \vote, with each acknowledgment sent to the immediate predecessor of the corresponding \vote. The acknowledgment payload, created by the \schemeacronym server (replacing $d$) is:
\begin{equation}
    r = Hash(d, s_{N_{shuffle}+1}, t_{timestamp}) \,,
    \label{equ:reply}
\end{equation}
where $s_{N_{shuffle}+1}$ is the symmetric key shared between the original sender and the \schemeacronym server. To reverse the path through the mix network, messages are sent as follows, where the receiving node encrypts the payload with $s_i$ instead of decrypting it:

\begin{enumerate}
    \item $Node_i$ receives a \vote packet containing $p_{i+1}$, $h_{i+1}$, and $r_{i+1}$ from $Node_{i+1}$ via the \schemeacronym server.
    \item $Node_i$ encrypts $r_{i+1}$ with $s_i$ to generate $r_i$.
    \item $Node_i$ replaces $p_{i+1}$ and $h_{i+1}$ with $p_i$ and $h_i$ (saved and used as flow identifiers from the original \vote shuffling rounds).
    \item $Node_i$ sends $p_i$, $h_i$, and $r_i$ to the \schemeacronym server, forwarding them to $Node_{i-1}$, which repeats the process (by tracking the \vote shuffling flows, the server can map $h_i$ to $Node_{i-1}$).
\end{enumerate}

When the original sender receives $r_1$ from $Node_1$, it can decrypt all $N_{shuffle}+1$ encryptions, as it possesses all symmetric keys. Subsequently, using Equation~\eqref{equ:reply}, it can authenticate that $r$ is the acknowledgment of the originally transmitted $d$. Similarly to the \schemeacronym server in Section~\ref{sec:scheme_next_hop_hash}, clients can trivially track when the acknowledgments should exit the \mixnetworkvotes.

In each acknowledgment shuffling round, each shuffling client expects a set of acknowledgments responding to the \votes it processed during the corresponding \vote shuffling round. Thus, the shuffling clients can generate dummy cover acknowledgments to replace any missing ones (potentially dropped by a malicious party) to ensure that no entity is able to drop the acknowledgment of a specific \vote and observe who does not receive it. The cover acknowledgments (generated in place of missing ones) are generated by appending random bytes to $p_i$ and $h_i$ in place of $r_i$. Thus, the original sender is notified of potential misbehavior as they could not authenticate the acknowledgment due to $r$ being random bytes.

\subsection{Resolving Records Outside the List}
\label{sec:scheme_external_resolutions}

Records not found in the \popularitylist are resolved externally using what we term a \emph{fallback DNS protocol}. The chosen fallback DNS protocol depends on the desired performance and privacy. We consider the following configurations, with (4) and (5) being the most relevant:

\begin{enumerate}
    \item \textbf{Unencrypted DNS:} No protection for queries outside the \popularitylist, but the overall exposure is reduced as most queries are resolved locally by the \popularitylist. We use Cloudflare, 1.0.0.1/1.1.1.1.
    \item \textbf{DoH:} Adds encryption to external queries to protect against external eavesdroppers. We use Cloudflare, 1.0.0.1/1.1.1.1.
    \item \textbf{DNSCrypt with rotating resolvers:} Continuously changing resolvers to reduce exposure to any individual one. We randomly query one of the DNSCrypt resolvers in \cite{relay_and_resolvers}.
    \item \textbf{Anonymized DNSCrypt with rotating relay/resolvers:} Reduces the exposure to any resolver-relay pair, where a large number of resolver-relay pairs can be rotated between \cite{relay_and_resolvers}. We randomly query one of the DNSCrypt resolvers in \cite{relay_and_resolvers}, each with a permanently assigned random relay.
    \item \textbf{DoHoT:} Lessens the risk of colluding relays and resolvers by forwarding queries through Tor, at the cost of high latency. We use a standard configuration of three nodes for Tor and Cloudflare's 1.0.0.1/1.1.1.1 as the resolver.
\end{enumerate}

\section{\schemeacronym Evaluation}
\label{sec:results}

In this section, we quantitatively evaluate \schemeacronym. We evaluate the hit ratio of the \popularitylist in Section~\ref{sec:results_hit_ratios}, the rate at which DNS queries are exposed to external entities in Section~\ref{sec:results_exposure_rate}, and the scheme's performance overhead, i.e., latency, memory/storage, and network bandwidth in Section~\ref{sec:results_latencies}, \ref{sec:results_memory_storage}, and \ref{sec:results_network_bandwidth}, respectively. We compare the results with widely deployed DNS privacy schemes. Due to space limitations, arguments for the security and unlinkability/anonymity of \schemeacronym are embedded in its presentation in Section~\ref{sec:scheme}, providing the reasoning behind each corresponding design choice.


\subsection{Dataset}

We simulate real-time operation using a dataset that contains real-world DNS queries and responses from a campus network with up to \num{4000} concurrent active hosts \cite{dataset}. Only DNS packets querying A or AAAA records (IPv4 and IPv6 addresses) are considered. Packets with malformed responses, server errors, and non-compliant FQDNs are excluded, as are retransmissions whenever they affect cache hit ratios. 172.31.3.121 serves as the main resolver for the network, indicated by its interaction with DNSSEC-related record types (used by resolvers to authenticate DNS data \cite{dnssec}). 172.31.3.121 communicates solely with 172.31.1.6, which serves thousands of clients. To avoid counting the same packet on multiple links or missing queries responded to by the DNS cache of an earlier router, only traffic between clients and 172.31.1.6 is considered.

\subsection{Popularity List Hit Ratio}
\label{sec:results_hit_ratios}

\subsubsection{Initialized List}

Figure~\ref{fig:hit_ratios} shows the measured hit ratio of the \popularitylist for different list sizes: the percentage of user queries in the dataset the list could resolve, as each DNS query is sequentially processed, simulating a real-world application of \schemeacronym. Shown is the mean hit ratio for each hour of the last nine out of ten days of the dataset, with each user (IP address) casting up to \num{10} \votes per hour ($t_{refresh} = \num{3600}$), with a voting rate of \num{0.3} (see Section~\ref{sec:scheme_tuning_parameters}). For the first \num{18} hours, all user queries are considered \votes to generate the initial \popularitylist (\textit{fast start} mode). The weight of the latest voting round (see Equation~\eqref{equ:weight}) is set to $a = \num{0.1}$. The values (default from here on) were selected to balance the hit ratio, privacy, and communication and computational overhead. The dashed plots show the mean hit ratio from day five and onward when no \votes other than from the first \num{18} hours have been contributed to the \popularitylist, simulating a sudden long-term stop in \votes.

\afterpage{
    \begin{figure}[!t]
        \centering
        \includesvg[width=0.9\linewidth]{images/results/hit_ratios.svg}
        \caption{The \popularitylist hit ratio during each hour of the day for various $N_{popular}$ values with $t_{refresh} = \num{3600}$, where the dashed plots simulate a long-term stop of all \votes.}
        \label{fig:hit_ratios}
    \end{figure}
}

The hit ratio quickly plateaus as the list size increases, achieving over \SI{90}{\percent} with $N_{popular} = \num{10000}$. Interestingly, $N_{popular} = \num{100}$ manages about \SI{40}{\percent}, demonstrating the high popularity of a small subset of DNS records. For optimal performance, a list size of at least \num{10000} is recommended, with \num{25000} considered the default value, achieving a mean daily hit ratio of around \SI{94.4}{\percent}, which means \SI{94.4}{\percent} of user DNS lookups are performed locally. Increasing the list size further to \num{100000} only marginally improves the hit ratio. The marginal drop in the hit ratio as no \votes are received after the first \num{18} hours, resulting in a \popularitylist that is between four and nine days out of date, shows that \schemeacronym is robust against Denial-of-Service (DoS) attacks on the voting system orchestrated by the \mixnetworkvotes (e.g., many clients dropping \votes).

\subsubsection{During Initialization with Few Users}

Figure~\ref{fig:low_participation} shows the mean hit ratio and mean number of records in the \popularitylist per day as it is instantiated with only a handful of participating users (during each voting round, only the \votes of a randomly selected subset of clients in the dataset is considered). Compared is with \textit{fast start} (users vote for all their performed DNS resolutions) and without.

\afterpage{
    \begin{figure}[!t]
        \centering
        \includesvg[width=0.9\linewidth]{images/results/low_participation.svg}
        \caption{Mean hit ratio (non-dashed) and mean $N_{popular}$ (dashed) per day for a \popularitylist initialized by only a few participating voting users.}
        \label{fig:low_participation}
    \end{figure}
}

The \popularitylist achieves a high hit ratio quickly even in cases where there are only a handful of users. The scheme achieves a mean hit ratio of around \SI{66}{\percent} during the second day with \votes from a maximum of ten users per hour (no \textit{fast start}). Around \SI{81}{\percent} is achieved if the number of voters increases to a mere \num{50}, matching that of \num{10} voters with \textit{fast start}. In other words, the initialization of the \popularitylist remains effective despite a minimal pool of participating users.

\subsection{Exposure/Loss of Privacy}
\label{sec:results_exposure_rate}


Figure~\ref{fig:exposure_rate} shows the \emph{exposure rate} of user DNS queries, i.e., the probability that a user-performed resolution can be connected to its user origin by a DNS resolver or the \schemeacronym server (due to encryption, these are the only respective entities with direct access to the content of queries). Compared are different protocols, considering varying collusion rates between anonymizing relays (relay in Anonymized DNSCrypt, Tor nodes, other clients in \schemeacronym, etc.) and the resolver/\schemeacronym server. Different fallback protocols and varying numbers of system users (voters) are compared for \schemeacronym while mixing \num{10} \votes in each of the \num{10} shuffling rounds for a list of size \num{25000}.

\afterpage{
    \begin{figure}[!t]
        \centering
        \includesvg[width=0.9\linewidth]{images/results/exposure_rate.svg}
        \caption{The probability that a DNS resolver or the \schemeacronym server can connect a user-performed DNS resolution to its IP address origin for different protocols and collusion rates between entities and the DNS resolver/\schemeacronym server. The DNS protocols on their own are plotted in blue, and are, when used as a fallback protocol for \schemeacronym using \num{50} and \num{10000} voters, plotted in orange and green, respectively.}
        \label{fig:exposure_rate}
    \end{figure}
}

Due to the lack of anonymizing relays, DoH offers no protection against curious DNS resolvers on its own, resulting in all queries being exposed. Anonymized DNSCrypt, with its single relay, can obscure user queries as long as the two entities are not colluding. DoHoT further reduces the exposure rate by using three separate relays. We have here assumed that no external observer (e.g., an ISP) performs volume or timing attacks on observed network traffic and that the pool of users (the anonymity set) is large enough that it can be approximated as infinite. \schemeacronym, without any such assumptions for its \mixnetworkvotes, significantly reduces the exposure rate in environments with colluding entities. The exposure rate for \schemeacronym includes that caused by \votes, as it is assumed that the number of voting users is small enough to affect privacy, hence \schemeacronym with \num{50} voters slightly increasing the exposure rate of Anonymized DNSCrypt and DoHoT as the collusion rate approaches zero in Figure~\ref{fig:exposure_rate}. Nevertheless, the exposure rate with \num{50} voting users is surprisingly similar to the, in practice, perfect setup of \num{10000}, implying that \schemeacronym is robust against poor user participation.

As expected, DoHoT as the fallback protocol results in the lowest exposure, with Anonymized DNSCrypt a close second. The generally low exposure rate of \schemeacronym is largely due to only around \SI{5.6}{\percent} of queries being resolved externally and around \SI{15.7}{\percent} being cast as \votes on average. The result is that \schemeacronym achieves an exposure rate of around \SI{20.5}{\percent} in scenarios where all nodes are colluding, providing the user with a higher guarantee of privacy compared to widely deployed alternatives. The exposure rate can be further reduced by decreasing the frequency of voting rounds (increasing $t_{refresh}$). The impact on the hit ratio should be minimal as indicated in Figure~\ref{fig:hit_ratios}, where a \popularitylist four to nine days out of date has a hit ratio only a few percentage points lower than one refreshed hourly.


\textbf{User Activity Status:}
The scheme reveals the activity status of users assigned to shuffling \votes by broadcasting their status to all users at the start of voting rounds. In addition, the server is trivially aware of which users are currently connected. However, specific activity levels (e.g., DNS queries per hour) are not revealed, mitigating most privacy concerns, as ISPs can obtain significantly more information regarding users' current activity rates.

\subsection{Lookup Latency}
\label{sec:results_latencies}

\begin{figure}[!t]
    \centering
    \includesvg[width=0.9\linewidth]{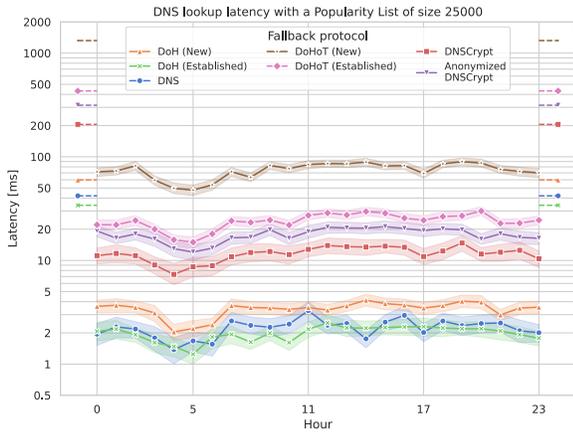}
    \caption{Mean \schemeacronym DNS lookup latency per hour using various fallback DNS protocols. Mean latency without \schemeacronym shown on the left and right ends (below \num{0} and above \num{23}).}
    \label{fig:latencies}
\end{figure}

Figure~\ref{fig:latencies} shows the mean DNS lookup latency of \schemeacronym, using a \popularitylist of size \num{25000} to resolve most DNS records. We compare the results of multiple different fallback DNS protocols to resolve queries/records outside the \popularitylist, as introduced in Section~\ref{sec:scheme_external_resolutions}. The mean latency of these DNS protocols on their own is displayed on the left and right sides of the figure to clearly show the significant improvements with the \popularitylist. Our latency measurements (measured using Wireshark/Tshark) were obtained by sampling \num{1000} lookup times based on the hit ratio from Figure~\ref{fig:hit_ratios}. For each sampled hit, a random record within the \popularitylist (implemented in C) is resolved in a UTM Kali Linux Virtual Machine (VM) on a Macbook Pro M1 Pro (four allocated CPU cores). For each miss, the same setup randomly queries one of the \num{500} most popular records in the dataset \cite{dataset} externally using a \SI{2.4}{\giga\hertz} Wi-Fi 6 connection, including the overhead of first querying the \popularitylist on the \emph{loopback} interface (but missing). Any domain in the 2016 dataset that is now an NXDOMAIN (non-existent domain) was not considered for the top \num{500}. In Figure~\ref{fig:latencies}, \quotes{New} includes opening a new TCP/TLS connection to the resolver, compared to \quotes{Established}.

Using a \popularitylist of size \num{25000} significantly decreases the mean latency across all DNS protocols. It is especially noticeable for new DoHoT connections, where the mean latency drops from over \SI{1}{\second} down to between \SI{80}{\milli\second} and \SI{90}{\milli\second}, greatly improving the viability of the protocol. However, one should note the variance in lookup latency for such a \popularitylist configuration, as most lookups are below \SI{1}{\milli\second} with a few exceeding \SI{1}{\second}.

\subsection{Memory/Storage Overhead}
\label{sec:results_memory_storage}

\subsubsection{Popularity List}

Figure~\ref{fig:bytes_cache} shows the mean size in bytes of the \popularitylist for varying values of $N_{popular}$ during all hours of the last nine days in the dataset \cite{dataset}. The impact of flattening CNAME chains (excluding the intermediate redirect domains) reveal modest but generally inessential gains. Overall, the \popularitylist scales well as $N_{popular}$ increases, where a \SI{150}{\percent} increase in size (\num{10000} to \num{25000}) results in only a \SI{60.5}{\percent} increase in byte size (no CNAME flattening). Considering that the main use case of the \schemeacronym scheme is to improve the privacy of web browsing, the memory overhead of the \popularitylist is negligible. For comparison, Google Chrome self-reports a memory usage of around \SI{30}{\mega\byte} to \SI{45}{\mega\byte} for a single tab of the Google Search landing page in Windows 11. Meanwhile, the YouTube homepage reported between \SI{250}{\mega\byte} and \SI{500}{\mega\byte}.

\afterpage{
    \begin{figure}[!t]
        \centering
        \subfloat[]{
            \label{fig:bytes_cache}
            \includesvg[width=0.45\linewidth]{images/results/cache_bytes.svg}
        }
        \hfill
        \subfloat[]{
            \label{fig:bytes_download}
            \includesvg[width=0.45\linewidth]{images/results/bytes_download.svg}
        }
        \caption{(a) \popularitylist size in bytes as a function of $N_{popular}$, with and without CNAME flattening. (b) Total sent and received bytes to download the \popularitylist (without CNAME flattening) over TLS, with and without compression.}
        \label{fig:bytes}
    \end{figure}
}

\subsubsection{Public Keys}

For a system with \num{10000} shuffling users, the storage overhead is \SI{0.32}{\mega\byte} (\SI{32}{\byte} per key), noticeably less than the \popularitylist of around \SI{1.2}{\mega\byte} (\num{25000} records without CNAME flattening in Figure~\ref{fig:bytes_cache}). The origins of these keys are verified through client certificates obtained during registration (see Section~\ref{sec:scheme}). We assume clients are not required to permanently store other clients' certificates locally. For example, the certificates might be publicly accessible on the \schemeacronym server, allowing clients to (optionally) fetch them to verify public keys at any point to reduce storage requirements.

\subsection{Network Bandwidth Overhead}
\label{sec:results_network_bandwidth}

All network bandwidth measurements are performed in an Ubuntu 22.04 VirtualBox VM, where all client-to-\schemeacronym server communications are sent over TLS using the Python \emph{ssl} library. Client-to-\schemeacronym server communications are simulated over the \emph{loopback} interface with a Maximum Transmission Unit (MTU) of \num{1500}.

\subsubsection{Downloading the Popularity List}

Figure~\ref{fig:bytes_download} shows the mean number of transmitted bytes to download the \popularitylist for each hour of the last nine days in the dataset \cite{dataset}. Compared is also the overhead when compressing the list using the Python \emph{zlib} library. Unsurprisingly, without compression, the number of bytes is similar to the size of the \popularitylist, as seen in Figure~\ref{fig:bytes_cache}. Compressing the \popularitylist significantly reduces the number of transmitted bytes, reducing it by around half.

\subsubsection{Continuous Client Traffic}

Figure~\ref{fig:bytes_total_client} shows the mean total hourly bandwidth overhead for each user/client (i.e., source IP address that generates at least one DNS query during the hour) in the dataset \cite{dataset} for various DNS configurations (excluding the initial download of the \popularitylist for \schemeacronym). The \schemeacronym configurations use Anonymized DNSCrypt with randomized resolver-relay pairs as its fallback protocol. Included for \schemeacronym is: (i) the overhead of receiving incremental updates to the \popularitylist (includes incremental updates to actively load-balanced records and replacing records due to changes in popularity, with compression applied), (ii) broadcasting the availability of shuffling clients, (iii) all \vote shuffling rounds, and (iv) falling back to Anonymized DNSCrypt according to the hit ratio in Figure~\ref{fig:hit_ratios}. The frequency that DNS answers change (causing the incremental updates) is extrapolated from measurements on the corresponding real-world domains by continuously re-querying a large subset of the domains in \cite{dataset} for an hour (only basing the measurements on domains that have not become NXDOMAINs since 2016). The bandwidth overhead of DNS queries is estimated using the mean overhead of resolving the top \num{500} most popular records from Section~\ref{sec:results_latencies}. The TCP-based protocols DoH and DoHoT use a \SI{30}{\second} connection keepalive.

\afterpage{
    \begin{figure}[!t]
        \centering
        \includesvg[width=0.9\linewidth]{images/results/bytes_client.svg}
        \caption{Total number of bytes sent and received per hour by a client for different DNS schemes, where \schemeacronym utilizes Anonymized DNSCrypt as its fallback protocol.}
        \label{fig:bytes_total_client}
    \end{figure}
}

The overhead of \schemeacronym using Anonymized DNSCrypt as fallback is similar to that of existing DNS protocols. Specifically, the default configuration (\num{25000} records, \num{10} \votes per user and hour, \num{10} shuffling rounds, \SI{60}{\second} minimum TTL for actively load-balanced records, and \num{10000} shuffling clients) is comparable to standalone Anonymized DNSCrypt, outperforming DoH (see (1) in Figure~\ref{fig:bytes_total_client}). With a minimum TTL of \SI{300}{\second}, \schemeacronym outperforms DNSCrypt (see (2)). A lighter configuration (\num{10000} records, \num{5} \votes, \num{3} shuffling rounds, \SI{30}{\minute} TTL, and \num{5000} mixing clients) matches the traditional DNS (see (3)) while improving privacy and reducing latency at the cost of less than \SI{1}{\mega\byte} (\num{10000} records and \num{5000} public keys) in memory/storage.

\subsubsection{Continuous Server Traffic}

Figure~\ref{fig:bytes_total_server} shows the mean network bandwidth usage of the \schemeacronym server for the scheme's default configuration for various $N_{popular}$. Included is the overhead of continuously re-querying public DNS resolvers using DoH to ensure records in the \popularitylist are up-to-date, and all data transmitted between the server and all its currently connected clients (excluding the initial client downloads of the \popularitylist). The overhead is considered insignificant, with only around \SI{25}{\mega\bit/\second} for the default configuration ($N_{popular} = \num{25000}$) with \num{50000} currently connected users. Even the most extreme configuration with $N_{popular} = \num{100000}$ and \num{100000} active users does not surpass \SI{100}{\mega\bit/\second}.

\afterpage{
    \begin{figure}[!t]
        \centering
        \includesvg[width=0.9\linewidth]{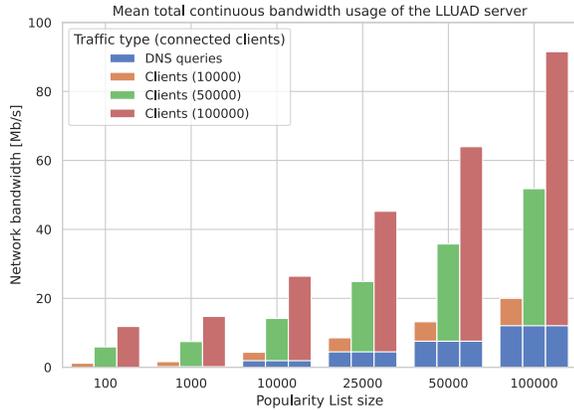}
        \caption{Mean continuous bandwidth usage of the \schemeacronym server, including: (i) re-querying DNS records in the \popularitylist as their TTLs expire, and (ii) all data continuously transmitted to/from connected users (\num{10} \votes per user, \num{10} shuffling rounds, \SI{60}{\second} minimum TTL, and \num{10000} mixing clients).}
        \label{fig:bytes_total_server}
    \end{figure}
}

\subsection{Comparison to Previous Work}

Table~\ref{tab:comparison_previous_work} compares \schemeacronym with the, to our knowledge, only conceptually directly comparable related work in \cite{dns_broadcast_range_queries_mix_zones}, where applicable. The hit ratio of \schemeacronym is generally in line with the \textit{Optimal TopList} from \cite{dns_broadcast_range_queries_mix_zones}, which is not implementable in real-world scenarios as it is constructed from future user queries. Although the hit ratio of \schemeacronym is slightly higher than the \textit{Optimal TopList} for medium-sized lists, the key takeaway is not the improvement itself, as much of it may be due to differences in the DNS dataset, but rather that \schemeacronym is comparable to an optimally instantiated \popularitylist.

\afterpage{
    \begin{table}[!t]
      \begin{center}
        \caption{Comparison between \schemeacronym and the results of \cite{dns_broadcast_range_queries_mix_zones} from 2011, where \textit{(comp.)} indicates whether compression is used.}
        \label{tab:comparison_previous_work}
        \begin{tabular}{|l|l|l|l|}
          \hline
          & \textbf{$N_{popular}$} & \textbf{\schemeacronym} & \textbf{\cite{dns_broadcast_range_queries_mix_zones}} \\ \hline
          \multirow{5}{6em}{\textbf{List mean hit ratio}} & \num{100} & \SI{39.2}{\percent} & \SI{40}{\percent} \\ \cline{2-4}
          & \num{1000} & \SI{72.6}{\percent} & \SI{63.9}{\percent} \\ \cline{2-4}
          & \num{10000} & \SI{91.4}{\percent} & \SI{83.9}{\percent} \\ \cline{2-4}
          & \num{25000} & \SI{94.4}{\percent} & \\ \cline{2-4}
          & \num{100000} & \SI{96.2}{\percent} & \SI{94.5}{\percent} \\ \hline
          \multirow{2}{6em}{\textbf{Downloading the list}} & \num{10000} & \SI{836.6}{\kilo\byte} & \SI{850}{\kilo\byte} \\ \cline{2-4}
          & \num{10000} (comp.) & \SI{426.1}{\kilo\byte} & \SI{290}{\kilo\byte} \\ \hline
          \multirow{3}{6em}{\textbf{Incremental updates}} & \num{10000} & \SI{74.1}{\kilo\byte/\hour} & \SI{2580}{\kilo\byte/\hour} \\ \cline{2-4}
          & \num{10000} (comp.) & \SI{67.5}{\kilo\byte/\hour} & \SI{1500}{\kilo\byte/\hour} \\ \cline{2-4}
          & \num{25000} (comp.) & \SI{139.9}{\kilo\byte/\hour} & \\ \hline
        \end{tabular}
      \end{center}
    \end{table}
}

For a list of size \num{10000}, \schemeacronym incurs a similar network overhead for downloading the \popularitylist without compression as \cite{dns_broadcast_range_queries_mix_zones} (\schemeacronym uses TLS, whereas \cite{dns_broadcast_range_queries_mix_zones} only uses TCP), suggesting comparable memory and storage requirements of the list, despite our \popularitylist including a \loadbalancingpool. The increased efficiency of our list is also shown by the comparably higher overhead with compression enabled. Although the \loadbalancingpool adds some memory/storage overhead to the list, it plays an important role in \schemeacronym reducing the network overhead of incremental updates. The improvement in network overhead of our \schemeacronym scheme is approximately \SI{95.5}{\percent} compared to \cite{dns_broadcast_range_queries_mix_zones} for $N_{popular} = \num{10000}$ with compression, where we use a negligible \SI{60}{\second} minimum TTL. These results are especially significant considering only \num{39} queries per second was required in \cite{dns_broadcast_range_queries_mix_zones} to continuously refresh the TTLs of \num{10000} records, compared to \num{188} today, indicating a significant decrease in average DNS record TTL and a potential increase in required incremental updates.

\section{Conclusion}
\label{sec:conclusion}

\schemeacronym using DoHoT achieves near zero-trust DNS privacy, guaranteeing a high level of DNS privacy despite large-scale collusion between anonymizing entities/relays. Moreover, the scheme greatly improves the mean latency of its paired DNS protocol, allowing DoHoT using Tor to be on par with traditional DNS over port \num{53}. We also find that \schemeacronym does not incur any additional network overhead to the paired DNS protocol, only introducing a modest memory/storage overhead of less than \SI{2}{MB}. \schemeacronym achieves this uncompromising combination of attributes, while remaining compatible with existing DNS infrastructure, by pre-fetching a local \popularitylist of the most frequently resolved records, securely instantiated and continually updated based on anonymous user votes. \schemeacronym contributes a \mixnetworkvotes{} -- a highly scalable self-sufficient mix network design, not requiring third-party hosted mix servers and achieving packet lengths independent of the number of traversed mix nodes: allowing users to anonymously and efficiently cast a centrally limited/enforced maximum number of votes. \schemeacronym also reduces the network bandwidth overhead required to incrementally update the \popularitylist by over \SI{95}{\percent} by contributing a \loadbalancingpool system.

\begin{acks}
This work was supported by the Swedish Research Council (VR).
\end{acks}

\bibliographystyle{ACM-Reference-Format}
\bibliography{references.bib}

\end{document}